\documentclass{article}

\usepackage{arxiv}
\usepackage{longtable}
\usepackage[utf8]{inputenc} 
\usepackage[T1]{fontenc}    
\usepackage{hyperref}       
\usepackage{url}            
\usepackage{booktabs}       
\usepackage{amsfonts}       
\usepackage{nicefrac}       
\usepackage{microtype}      
\usepackage{lipsum}		
\usepackage{graphicx}
\usepackage{natbib}
\usepackage{doi}

\title{Boardwalk Empire: How Generative AI is Revolutionizing Economic Paradigms}

\author{Subramanyam Sahoo\thanks{Subramanyam Sahoo conducted this work during his Master's program at the NIT Hamirpur, under Kamlesh Dutta.
for correspondance shoot an email to \textbf{sahoosubramanyam@gmail.com}} \\
	Department of Computer Science\\
	National Institute of Technology, Hamirpur\\
	Himachal Pradesh , 177005 \\
	\texttt{22mcs107@nith.ac.in} \\
	\And
	Kamlesh dutta \\
	Department of Computer Science\\
	National Institute of Technology, Hamirpur\\
	Himachal Pradesh , 177005 \\
	\texttt{kd@nith.ac.in} \\
}

\hypersetup{
pdftitle={Boardwalk Empire: How Generative AI is Revolutionizing Economic Paradigms},
pdfsubject={Machine Learning},
pdfauthor={Subramanyam Sahoo, Kamlesh Dutta},
pdfkeywords={Deep generative models , Large language Models , Finance Variable},
}
\begin{document}
\maketitle

\begin{abstract}
	
The relentless pursuit of technological advancements has ushered in a new era where artificial intelligence (AI) is not only a powerful tool but also a critical economic driver. At the forefront of this transformation is Generative AI, which is catalyzing a paradigm shift across industries. Deep generative models, an integration of generative and deep learning techniques, excel in creating new data beyond analyzing existing ones, revolutionizing sectors from production and manufacturing to finance. By automating design, optimization, and innovation cycles- Generative AI is reshaping core industrial processes. In the financial sector, it is transforming risk assessment, trading strategies, and forecasting, demonstrating its profound impact. This paper explores the sweeping changes driven by deep learning models like Large Language Models (LLMs), highlighting their potential to foster innovative business models, disruptive technologies, and novel economic landscapes.

\end{abstract}


\section{Introduction}

Since its beginnings, \textbf{artificial intelligence} has been the talk of the twenty-first century. Its ability to alter the dynamics of daily life has made it popular among blue-collar workers, researchers, business domain experts, and neuro-scientists. Different ways have been utilized to define AI, and in a majority of situations, \textbf{Machine Learning} (Statistical methods) principles have been applied. But, unlike traditional statistical methods, the new paradigm called \textbf{Deep learning} has been on the rise due to its ability to compute enormous amounts of multidimensional datasets. Deep learning configures prediction-based capabilities that rival or surpass human intelligence in a variety of ways, which is further supported by the fact that it can perform a very qualitative analysis of complex hierarchical data representations that a human mind may find difficult \cite{nalisnick2018deep}. Thus, there is a good chance that it will be used in business domain problems. This decade has seen the inception of a new framework known as Artificial General Intelligence(AGI). Elements such as \textit{ChatGPT} and \textit{Midjourney} have shifted the world in a new direction. In this present paper, authors attempt to show a futuristic strategy to solve finance related problems using data-driven Generative AI models.

\begin{figure}
    \centering
    \includegraphics[width=0.75\linewidth]{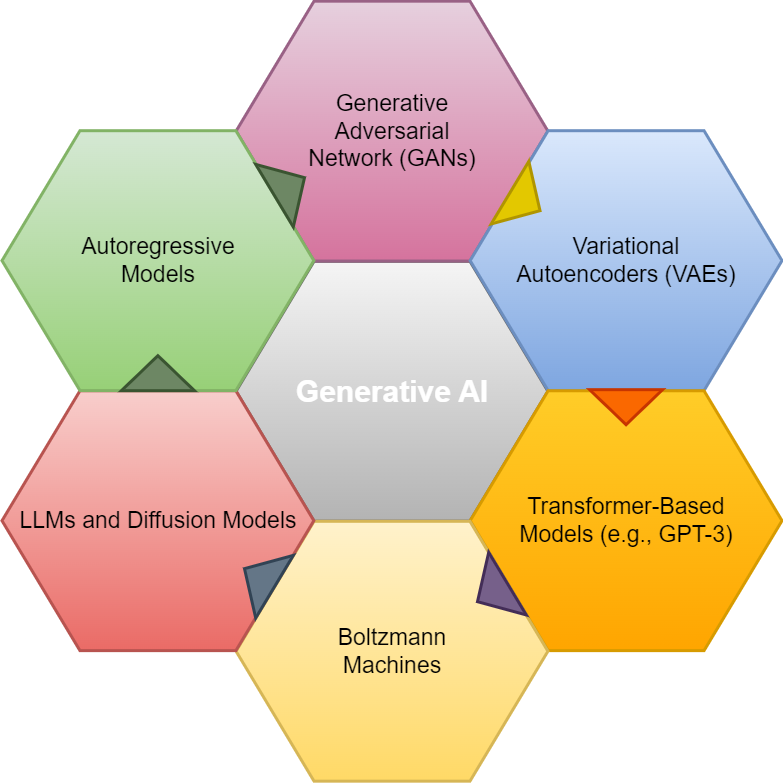}
    \caption{Different Generative AI Frameworks}
    \label{fig: Different Generative AI Frameworks}
\end{figure}

Deep generative models are enthralling and crucial models of artificial intelligence, notably in machine learning. These models are intended to produce new data samples that seem similar to those in an existing dataset. They have risen to prominence due to their ability to generate realistic data in a variety of areas, including images, text, and others. Deep generative models are utilized in picture synthesis, text generation, data denoising, and anomaly detection, among other things. Through a sequence of invertible transformations, these models focus on changing a simple probability distribution into a more complex one. To represent complex, multi-modal data distributions, normalizing flows are used\cite{salakhutdinov2015learning}. Since the inception of ChatGPT by OpenAI, Generative AI has been the buzzword. Deep Generative models' great power has been unleashed on humanity to address resource-limited complex issues with a substantially lower time complexity\cite{kalla2023study}. Generative AI's major goal is to generate data that is indistinguishable from real data. It is a sort of model that can generate new content in terms of writing, images, and music. Most of this kind of power comes from the family of deep unsupervised learning. Deep unsupervised methods employ complex mechanisms to extract information from the latent representation of data. This family of algorithms can learn complicated patterns, allowing them to create more realistic and imaginative content. It is a fast-developing field that blends unsupervised learning techniques with deep learning models to construct generative products. These systems are capable of producing fresh and meaningful data. Numerous methods, including Autoregressive models, Variational Autoencoders (VAEs), and Generative Adversarial Networks (GANs), are part of deep generative AI \cite{wang2023generative}. Finding and comprehending the underlying structures and patterns in the data is the aim of these models. These structural approaches enable them to generate new samples that closely mimic the original data distribution. In domains including computer vision, natural language processing, medication discovery, art, product design, financial forecasting, and music production, to mention a few, the capacity to produce realistic data has far-reaching implications.

\begin{figure}
    \centering
    \includegraphics[width=1\linewidth]{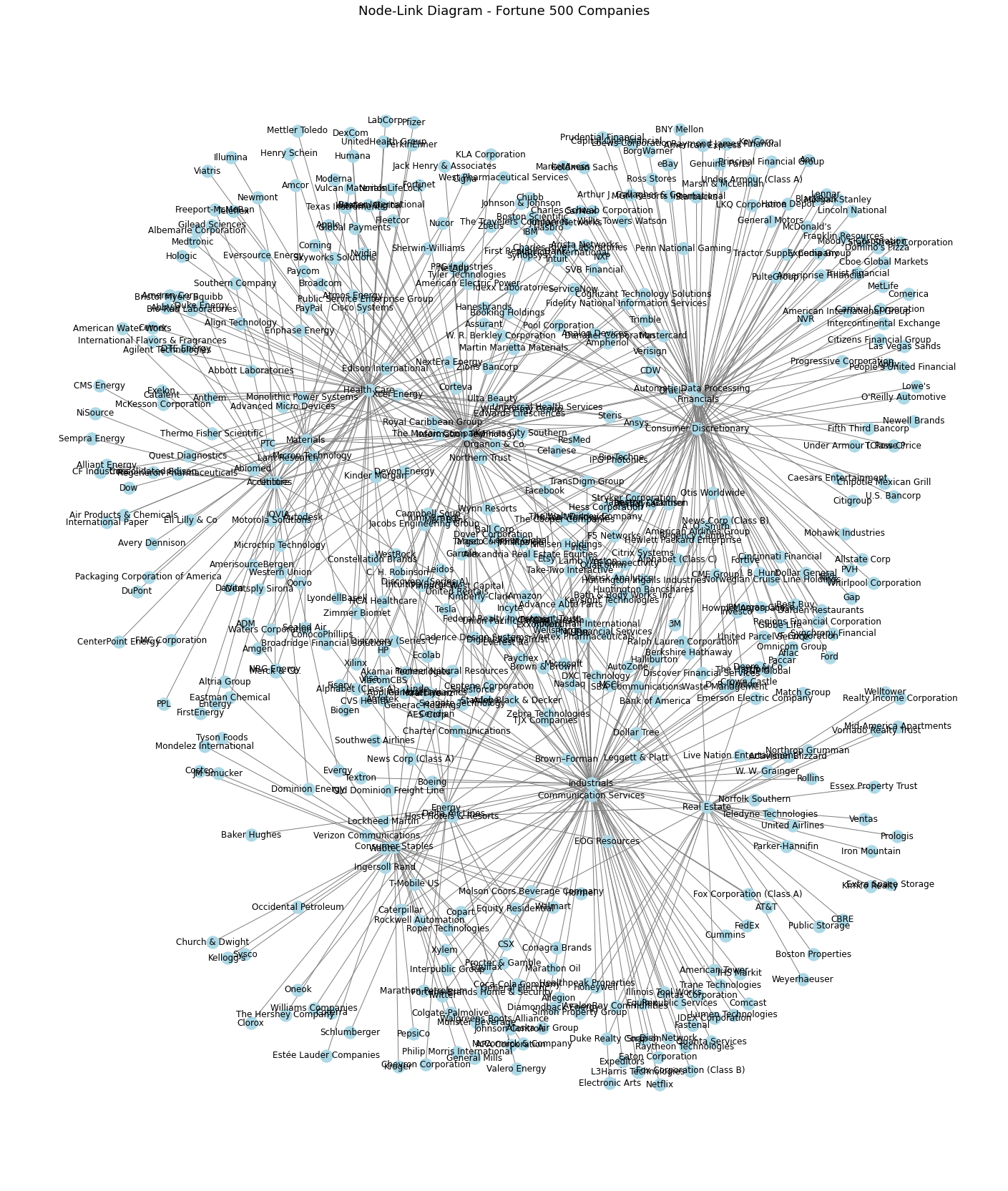}
    \caption{\textbf{Node - Link diagram for Fortune 500 companies}}
    \label{fig: Node - Link diagram for Fortune 500 companies}
\end{figure}

\begin{figure}
    \centering
    \includegraphics[width=1\linewidth]{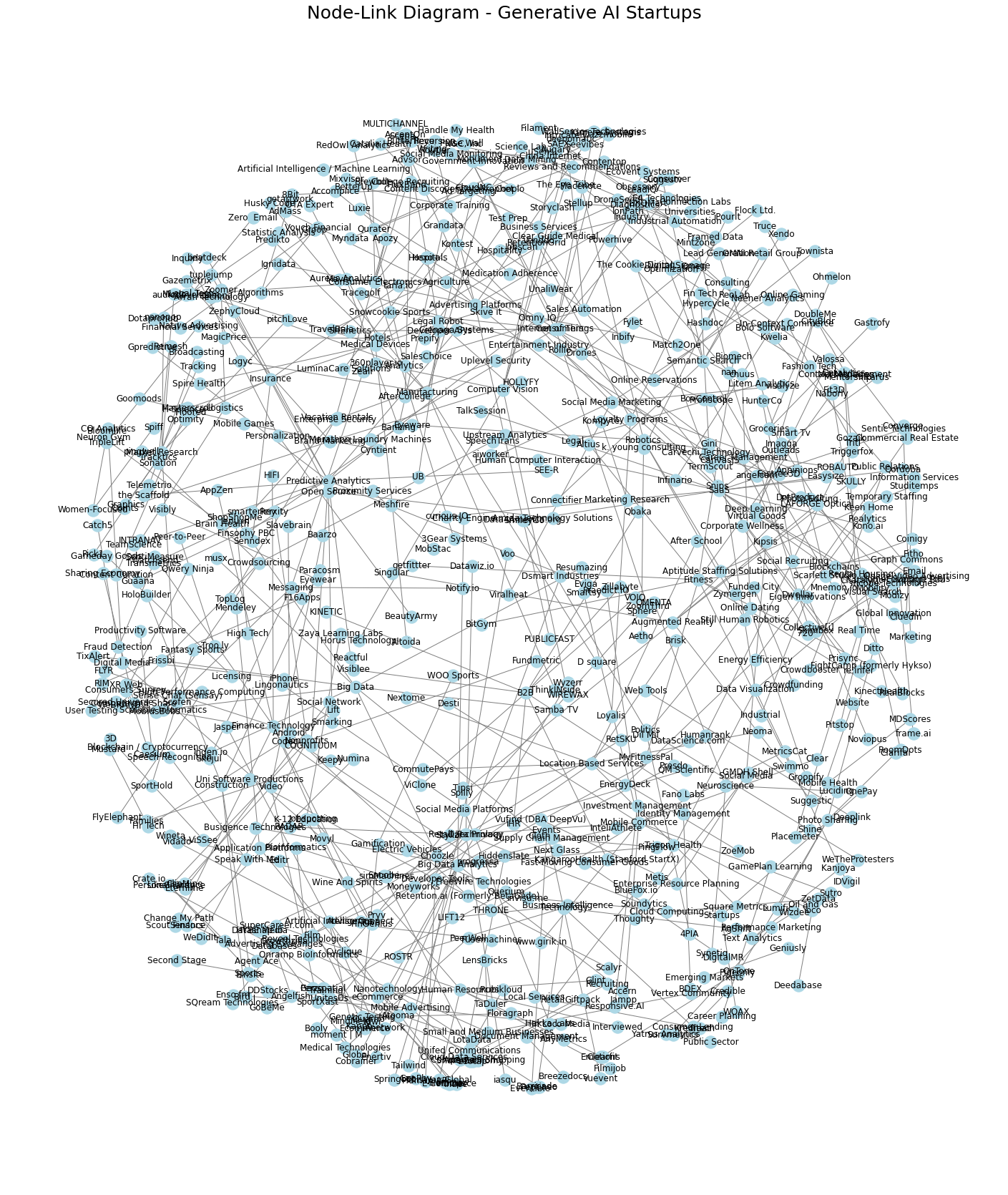}
    \caption{\textbf{Node - Link diagram for Generative AI startups}}
    \label{fig: Node - Link diagram for Generative AI startups}
\end{figure}

These models have the potential to revolutionize creative industries\cite{li2023chatgpt} by assisting in data augmentation, improving simulation settings, and aiding in the interpretation of complex data distributions. Table \ref{tab:my-table Audio  2 Everything} shows generative models where audio data is taken as a source and Table \ref{tab:my-table Text  2 Everything} shows generative models where text data is taken as a source respectively.
This paper scores some quick ideas about how these AI models will attempt to leave their mark throughout time. The concept of how specific company domains might achieve exponential growth by utilizing the aforementioned frameworks along with their limitations shall be discussed in this paper.  Potential weaknesses have been noted, as have potential opportunities. A network of Fortune 500 firms as shown in Fig  \ref{fig: Node - Link diagram for Fortune 500 companies} is also provided as a \textbf{Node-Link} graph.  The majority of these organizations have the means to access a deployable Gen AI model for further market expansion. Another \textbf{Node - Link} diagram is shown to visualize the importance of generative analytics for startups and big tech companies in Fig~\ref{fig: Node - Link diagram for Generative AI startups}. Through real-world solutions, the article conceptualizes the innovative idea of futuristic technologies. Attempts to broaden the research domains by providing an insight into existing research gaps and how Generative AI will transform society while minimizing the distance between cyberspace and real space shall also be explored. The crux of the study will be how the practical application of the aforementioned methods will have a positive impact on society\cite{ye2023financial}. The main focus in this paper is  on how finance-related services are going to be transformed by using AI models.

The remaining paper is organized as follows. In sections 2 and 3, various aspects of methodology and why Generative AI is going to be good for business is discussed. In section 4, different kinds of generative foundation models have been described. In section 5,  opportunities in finance are  explored through different working mechanisms. In section 6 and 7, real-world solutions along with the effect of generative AI are  discussed. In section 8, the limitations of these models is described. In section 9, suggestions are  provided for entrepreneurs and startups. Some possible future research directions are also listed. The last section summarises the observations with concluding remarks.

\section{Economic and Financial Variables}\label{sec2}

Economic variables are indicators of the state of the economy at the moment. To comprehend the factors influencing economic growth, the timing and mechanisms of price increases, the causes of inflation, and the best kinds of state-appropriate measures. Macroeconomic performance (gross domestic product [GDP], investment, trade, and consumption) and stability (central government budgets, prices, money supply, and balance of payments) are measured by economic indicators\cite{cao2020ai}. The main reason why financial variables are used to forecast future economic events is that they are the best representations of investors' and other economic agents' expectations and behavior. While economics takes into account both material and non-material resources and how resource scarcity may affect local or global markets, commodities and services, and human behavior, finance is defined in many respects by the actual usages of money\cite{terasvirta2006forecasting}. Primarily, finance relates to the management of money. As we all know, disruption is the secret sauce to capitalism, and \textit{change is the only constant}. So how can AI exist without having an impact on people's lives, both implicitly and explicitly!!! Recent developments demonstrate that multimodal and multidimensional Generative AI paradigms are increasingly emerging, raising concerns about their economic implications. The direction of both micro and macroeconomic variables will be influenced by factors such as established enterprises and startups utilizing AI. It, like previous innovations, will alter the way people work and play. The myth and cult around AGI - Artificial General Intelligence \cite{brand2023using} among researchers, managers, financial advisors, CEOs, and investors is considerable, and the Industry 4.0 structure is heavily reliant on these frameworks. Companies are aiming for products with AI-integrated functionalities to obtain a significant market value. Product-based businesses that successfully integrate AI as a service into their offerings are likely to gain a competitive advantage, particularly those who can fine-tune their models with unique and useful data sets for specific use cases.

The nexus between Generative AI (GenAI) and financial and economic variables in the current landscape of artificial intelligence (AI) applications represents a paradigm shift in the modeling, analysis, and prediction of economic occurrences. The complex interplay between GenAI and financial and economic variables presents a diverse range of opportunities and difficulties, hence transforming the traditional approaches utilized in financial and economic modelling\cite{aghion2017artificial}.
The modeling of economic variables is one important area where the effects of generative AI are seen. The economic variables that fall under the purview of Generative AI encompass a wide range of factors, ranging from microeconomic dynamics to macroeconomic indices. Generative AI models are used to simulate and analyze the effects of market movements, economic policies, and the dynamics of international commerce at a macro level. These models can mimic different economic situations, giving policymakers insight into possible outcomes and supporting them in making well-informed decisions. Conventional economic models frequently make assumptions that might not fully represent the intricacy of actual economic systems. With the use of GenAI, data-driven methodology can be used to create artificial economic variables that closely resemble observable trends. This improves economic models' accuracy and advances our knowledge of the underlying processes that control economic indicators like GDP, inflation, and employment rates. The application of GenAI to financial variables is a major step in the field of finance. The innate volatility and non-linearity of financial markets pose a challenge to financial models\cite{arslanian2019future}. Artificial intelligence (AI) generative approaches make it easier to create synthetic financial data, which in turn makes it possible to create prediction models that capture the complex correlations between variables such as asset returns, market volatility, and stock prices. The capacity to produce a variety of financial scenarios improves risk management tactics and helps make more sound investment selections. Generative AI is used for more than just modeling when combined with financial and economic factors. It gives decision-makers a useful tool for scenario analysis and strategy planning by offering a unique capacity for modeling a variety of economic and financial scenarios. Decision-makers can evaluate the resilience of financial and economic systems under different conditions with the help of GenAI, which produces believable but previously unseen data points7\cite{packin2019consumer}. This leads to better-informed decision-making processes. Advanced mathematical models, machine learning techniques, and statistical analyses are applied in the integration of economic and financial data with Generative AI. Several scientific approaches are utilized to capture the intricate interdependencies found in economic and financial systems, including time-series forecasting, probabilistic modeling, and Monte Carlo simulations.

\section{Rise of Deep Generative Models}\label{sec4}
GenAI represents a significant advancement in AI technology, increasing its utility for financial institutions that have been quick to apply it to a wide range of applications. However, there are hazards inherent in AI technology and its implementation in the financial sector, such as embedded bias, privacy concerns, outcome opacity, performance robustness, unique cyber threats, and the possibility for new sources and transmission channels of systemic problems\cite{ye2023financial}. GenAI could exacerbate some of these issues while also introducing new forms of vulnerabilities, particularly to financial sector stability. This study offers preliminary insights into the inherent dangers and solutions of GenAI and their potential influence on the financial sector. The adoption of generative AI in banking functions, like that of other technologies, will most likely follow an Exponential pattern. Finance teams are currently investigating how technology may supplement conventional procedures by writing text conducting research and creating preliminary draughts for jobs that are text-heavy or require little analysis, such as contract drafting and credit review supplementation. Accounting and financial reporting are two terms used interchangeably\cite{houde2020business}. Providing preliminary insights to assist in later versions of financial statements during month-end closes, or assisting in the construction of audit trails for reclassification memorandum. Budgeting and performance management are critical components of financial planning. Conducting ad hoc variance studies using the company's structured or unstructured data sets, such as comparing actual to forecasts, and producing reports to explain the financial performance of various business units to stakeholders. The adoption of AI solutions in the banking sector will be accelerated by the market growth rate. Competitive forces have fuelled the financial sector's rapid adoption of machine Learning in recent years by facilitating gains in efficiency and cost reductions, redefining client interfaces, improving forecasting accuracy, and improving risk management and compliance\cite{strasser2023pitfalls}.In the financial sector, generative models have proven useful for some useful tasks as these organizations can improve their ability to foresee market trends and simulate various economic situations by using generative models, allowing them to make more educated investment decisions. Some of the used models are discussed in this section Fig~\ref{fig: Different Generative AI Frameworks}.
\subsection{Variational Autoencoders (VAEs)}\label{sec4.1}

 These are generative AI models frequently employed in the banking industry. VAEs\cite{kingma2019introduction} are intended to understand the underlying structure of the input data and generate new samples that are similar to the original data distribution. Within the financial domain, these models operate through a process of dimensional reduction, whereby input financial data is mapped into a lower-dimensional latent space representation. This latent encoding adeptly captures the intrinsic attributes and latent patterns inherent to the dataset. The encoded data is subsequently subjected to a reverse transformation, enabling a reconstitution of the original data in its native space, thereby facilitating the restoration of the initial input data. The training entails optimizing two goals: reconstruction loss and Kullback-Leibler (KL) \cite{bu2018estimation}divergence. The reconstruction loss serves as a metric to assess the dissonance between the input and reconstructed data, thus incentivizing the model to furnish precise representations. Simultaneously, the KL divergence imposes constraints upon the latent space, coercing it to conform to a predefined distribution, typically a standard normal distribution, thereby imparting a regularization effect. This regularisation encourages the creation of varied and meaningful samples. VAEs are used in a \cite{xu2019explainable} variety of sectors in finance, including Portfolio optimization in which using historical market data, VAEs may learn the underlying structure and build new investment portfolios. Then in  Anomaly detection VAEs are capable of detecting abnormal patterns in financial transactions or market behaviour. Risk assessment and modeling are another kind of field in which an Autoencoder can be used to simulate and assess hazards. Sophisticated CrossConvolutional Autoencoder can aid in the detection of fraudulent financial transactions. Variational Autoencoders (VAEs) exhibit the capability to synthesize financial data synthetically, thereby addressing limitations inherent in authentic real-world datasets. Notably, they find extensive application within options trading, where they aid in the creation of synthetic volatility surfaces, thereby augmenting the precision of option pricing and affording refined methodologies for trading and risk evaluation.
 
\subsection{Generative Adversarial Networks (GANs)}

In finance, generative adversarial networks (GANs)\cite{goodfellow2016nips} are utilized for tasks such as synthetic data generation, market simulation, and risk modeling improvement. These are generative AI models that have two components: a generator and a discriminator. They have acquired enormous attraction and adoption within the sphere of finance due to their ability to generate synthetic data and boost various financial procedures. The training procedure takes the form of an enthralling antagonistic dance between the generator and the discriminator. The generator's goal is to deceive the discriminator by providing specimens that gradually resemble legitimate data, whereas the discriminator's goal is to improve its ability to distinguish between the genuine and the fake. As training progresses, the generator learns to provide progressively very similar financial data, while the discriminator hones its expertise in the fine art of distinguishing between veracity and deception.

GAN applications in finance include that it can produce synthetic financial data, and addressing difficulties such as limited or biased datasets. Risk modeling, algorithmic trading, and portfolio optimization can all benefit from this information. Financial fraud detection: GANs can help distinguish between real and fraudulent transactions, improving financial fraud detection.GANs can generate artificial market data, assisting in understanding market dynamics, anticipating price changes, and evaluating the impact of various factors on financial markets. GANs can find anomalies in financial data by detecting unexpected patterns or outliers\cite{ngwenduna2021alleviating}. CTAB-GAN\cite{zhao2021ctab}, a conditional GAN-based tabular data generator, to produce synthetic data for credit card transactions, outperforms earlier methods. To detect fraud in imbalanced credit card transactions, a new model named Generative Adversarial Fusion Network (IGAFN)\cite{lei2020generative} is used. Integration of diverse credit data was expertly choreographed, alleviating the problem of data asymmetry and exhibiting superior prowess when compared to existing credit scoring algorithms. These findings eloquently demonstrate the effectiveness of Generative Adversarial Networks (GANs) in uncovering the threat of credit card theft, with the tantalizing promise of ushering in advances in risk assessment within the financial arena.

\subsection{Autoregressive Models}
Temporal chronomancers, also known as time series models, rule supreme in finance, providing important instruments for diving into the complexities of analysis and prediction. These sophisticated models can decipher the delicate threads of temporal links and patterns within sequential data, whether it's the cryptic undulations of stock prices, the rhythmic cadence of interest rates, or the ebbs and flows of economic indicators.

 Autoregressive models\cite{bond2103deep} operate on the premise that the value of a variable at a given point in time is determined by its prior values. High dimensional models, such as autoregressive moving average (ARMA) and autoregressive integrated moving average (ARIMA), analyze the relationship between an observation and a lag set of observations. At its essence, the fundamental concept revolves around the notion that the state of a variable in a specific moment can be foretold by summoning forth a symphony of its past incarnations, garnished with a dash of stochastic serendipity. The term autoregressive elegantly alludes to this dynamic reliance on antecedent renditions of the variable. This model adroitly assigns significance to these temporal echoes, orchestrating their harmonious influence to orchestrate a compelling forecast of the present state \cite{doering2019metaheuristics}. In the case of ARMA models, the moving average element alludes to the model's reliance on previous forecast errors or residuals. These models are often estimated using historical data to minimize the discrepancy between the observed and anticipated values.

The financial deployment of autoregressive models creates a landscape brimming with promise, where future financial variables are created from echoes of their past. These models summon their prophetic prowess in this mystical realm to foretell the dance of stock market prices, the flux of interest rates, the enigmatic waltz of currency exchange rates, and the harmonious rhythms of various financial indices, creating a symphony of insight into the financial future. These models help in risk assessment and portfolio optimization by modeling the volatility and correlations of asset returns. On the other hand, models assume stationarity, which means that the statistical features\cite{mohapatra2020evaluation} of the data remain constant throughout time. As a result, it is critical to evaluate the data's stationarity and, if necessary, apply transformations or investigate more advanced models, such as ARIMA, which integrates differences to solve non-stationarity.

\subsection{Transformers}
A transformer\cite{vaswani2017attention} is a sort of neural network architecture that has gained popularity due to its capacity to handle sequential input more efficiently, such as text. Since transformer models can understand long-term dependencies and deal with the complex maze of sequential data, they have gained an important place in the financial domain. Their uses are numerous and include sentiment analysis, orchestration of document classification, and creation of financial text compositions. At the core of a transformer model is the attention mechanism, a vital component in the computational gear. This technique effectively allows the model the ability to generate representations while distributing different weights or significance throughout the input sequence's structure. It turns into a tool for the model to focus its perceptive attention on pertinent sections, reliably capturing the subtleties of inter-element relationships.
Transformer models, with their exceptional ability, untangle the intricate emotional fabric inherent in financial news, social media communications, and other textual transmissions. They expertly gather contextual cues and analyze word inter-dependencies, revealing illuminating insights into market emotion and providing investors with discernible tools for educated decision-making.
Transformer models have a wide range of applications, including document classification. They take on the task of categorizing a variety of financial documents, research articles, and other textual expositions, producing a symphony of organization amid the textual cacophony.
 \cite{lopez2022multi}. This aids in the organization and filtering of enormous amounts of financial data. Transformer models act as alchemists in the world of textual finance, conjuring up fictitious financial treatises, market exegesis, and relevant literary creations. Their art is a data-driven linguistic creative dance, informed by the intricate patterns and structures hidden inside the immense tapestry of financial data. This command of the language allows for the orchestration of automated report production and the seamless generation of content, ushering in a new era of textual creation built from algorithmic enchantments.

\begin{table}[]
\centering
\resizebox{\columnwidth}{!}{%
\begin{tabular}{|c|c|c|c|c|}
\hline
\textbf{Models} &
  \textbf{\begin{tabular}[c]{@{}c@{}}Restricted \\ Boltzmann \\ Machine (RBM)\end{tabular}} &
  \textbf{\begin{tabular}[c]{@{}c@{}}Variational \\ Autoencoder\\ (VAE)\end{tabular}} &
  \textbf{\begin{tabular}[c]{@{}c@{}}Autoregressive \\ Models\\ (LSTM , Transformers)\end{tabular}} &
  \textbf{\begin{tabular}[c]{@{}c@{}}Generative \\ Adversarial \\ Network (GAN)\end{tabular}} \\ \hline
\textit{\textbf{Abstraction}} &
  Yes &
  Yes &
  No &
  No \\ \hline
\textit{\textbf{Generation}} &
  Yes &
  Yes &
  Yes &
  Yes \\ \hline
\textit{\textbf{\begin{tabular}[c]{@{}c@{}}Probability\\ Computation\end{tabular}}} &
  Intractible &
  Intractible &
  Tractible &
  No \\ \hline
\textit{\textbf{\begin{tabular}[c]{@{}c@{}}Sampling\\ Speed\end{tabular}}} &
  \begin{tabular}[c]{@{}c@{}}Markov Chain\\ Monte Carlo\end{tabular} &
  Fast &
  Slow &
  Fast \\ \hline
\textit{\textbf{\begin{tabular}[c]{@{}c@{}}Types of\\ Graphical\\ Models\end{tabular}}} &
  Undirected &
  Directed &
  Directed &
  Directed \\ \hline
\textit{\textbf{\begin{tabular}[c]{@{}c@{}}Loss \\ Function\end{tabular}}} &
  KL divergence &
  KL divergence &
  KL divergence &
  JS divergence \\ \hline
\textit{\textbf{Samples}} &
  Very Bad &
  OK &
  Good &
  Best \\ \hline
\end{tabular}%
}
\caption{A Comparison of Generative Models based on Architecture}
\label{tab:A Comparison of Generative Models based on Architecture}
\end{table}

Table \ref{tab:A Comparison of Generative Models based on Architecture} shows the power of different generative models and it is suggested that while creating models, these aspects should be taken into consideration to maintain the sanity of output products.

\section{Opportunities in Finance}\label{sec5}
Generative AI opens a wide range of opportunities in the finance domain. Generative models explore different paradigms of structural solutions to some of the hardest solutions in the finance sector. The banking industry may improve efficiency, streamline decision-making processes, and better meet changing customer and regulatory requirements by implementing Generative AI in these areas. This will ultimately contribute to a more robust and inclusive financial system for everyone.

\subsection{Fraud detection and prevention}
The detection and prevention of fraud are key issues for the banking and financial services industries. The ever-changing fraudulent operations present substantial hurdles for institutions seeking to protect their systems and clients. Traditional rule-based systems and static models frequently fail to keep up with the complex strategies used by fraudsters. Furthermore, the enormous number of server-based transactions and supplementary data generated makes it impossible to detect fraudulent trends manually and quickly. This involves the investigation of advanced technologies such as generative AI to improve fraud detection and prevention capabilities\cite{niu2019comparison}. Generative AI efficiently synthesizes data with patterns that look fake. Synthetic data that mimics the features of fraudulent activities can be generated by models that have been trained on large datasets that include fraudulent examples. Financial institutions can use generated data to test and optimize realistic systems. Introducing a wider range of potentially fraudulent activities to these algorithms improves the institution's ability to identify and prevent fraud. This improves the institution's ability to repel expert con artists by enabling the creation of more resilient algorithms that can adapt to shifting fraud strategies\cite{chen2018credit}. By using artificial intelligence (AI) models for training, a wider range of bizarre behaviors can be taught to them. The models' capacity for prediction is improved by this addition. Gen AI offers several advantages for financial transaction security. Financial institutions can use generative artificial intelligence to proactively detect and prevent fraudulent activities. In this approach, customer accounts and assets are safeguarded. Organizations can test their fraud detection systems and replicate fraudulent tendencies by creating synthetic data. This iterative procedure enhances the system's dependability and efficiency over time. Ultimately, this boosts customer trust in the organization's security procedures
\cite{zheng2018generative}.

\subsection{Client Relationship}

Financial and banking services must offer individualized experiences to their clients. Today's customers demand solutions that are tailored to their specific needs and preferences. Financial institutions can increase customer engagement, create stronger ties, and stand out in a crowded market by offering personalized experiences. Financial service providers can win their client's trust and loyalty by being empathetic and showing gratitude in their offers\cite{micu2022assessing}. Generative AI generates customized financial advice based on specific consumer data. To provide individualized recommendations, very intelligent algorithms comb through a vast amount of customer data, including financial objectives and transaction history.
Customers are better prepared to make decisions about investing, budgeting, saving, and their overall financial well-being because of this customized counsel. Using information unique to each customer, such as their investing objectives and risk tolerance, generative AI algorithms create customized investment portfolios. To provide customers with investment recommendations that align with their financial goals, asset allocation is enhanced through the use of sophisticated algorithms and historical market data\cite{goldenberg2021personalization}. Furthermore, by taking into account previous transactions, customer behavior, and preferences, generative AI expands bespoke offerings to include product recommendations. These recommendations, which cover credit cards, insurance, loans, and investment products, raise consumer satisfaction and conversion rates. Financial institutions that take advantage of upselling\cite{parise2016solving}  and cross-selling opportunities can increase revenue and client lifetime value.

\subsection{Risk assessment and Credit scoring}

Financial organizations take into account the risks involved in granting credit to borrowers in addition to evaluating their creditworthiness. Conventional systems may underestimate the complexity of credit difficulties because they depend too heavily on past data and preset norms. Financial data and credit history are used to calculate credit ratings. With cutting-edge methods, generative AI improves risk assessment and credit scoring while producing synthetic data for model training. For efficient model training, generative AI manipulates synthetic datasets with a variety of risk situations\cite{solaiman2023gradient}. As a result, learning is enhanced and risk assessments are more precise. Algorithms comb through customer data, including bank account records and payment histories, looking for trends that indicate a person's dependability. Banks can make informed decisions about loans, interest rates, and credit limits by utilizing controlled methods that provide insights. The use of generative AI in credit rating systems improves banking risk management procedures. This partnership lowers default rates by making loan decisions with greater precision, dependability, and timeliness \cite{weisz2023toward}. Organizations can enhance the overall efficacy of risk management by employing modeling artistry, which facilitates scenario simulation, element-by-element risk analysis, and risk anticipation and navigation.

Capital allocation finds optimal resonance in this orchestration, losses are held at bay, and a harmonious risk-to-reward ratio is maintained. As the skilled conductor of automation, generative AI graces the stage to bring efficiency into the vast opera of risk management. It reveals doors for streamlined risk assessment procedures, the reduction of temporal turnarounds, and the acceleration of decision-making tempo. Institutions are becoming attuned to this modern symphony of algorithms and synthetic data, synchronizing their operations with the cadence of advancement and proficiency.
This enables institutions to handle higher volumes of risk assessments while maintaining accuracy and quality\cite{han2021explainable}. Financial institutions can use generative AI to simulate scenarios and analyze risk elements in a controlled environment. Institutions can examine the possible impact of numerous events on their portfolios and overall risk exposure by generating synthetic data modules that represent multiple risk scenarios. Banks can use Deep complex models to identify correlations, relationships, and developing dangers that standard risk assessment approaches may miss \cite{kang2022cwgan}. This proactive strategy assists institutions in developing robust risk management strategies and making educated risk-mitigation decisions.

\subsection{Chatbots and Virtual Assistants}

Virtual assistants have gained significant traction in the banking and financial services industry as tools to enhance customer support and engagement\cite{radford2023robust}. AI-driven conversational agents emerge as natural language engagement experts in the expanding world of digital interactions, orchestrating smooth conversations with customers and ushering in the era of automated support and query resolution. Their constant presence, available at all hours of the day and night, provides an unbroken line of communication for customers, epitomizing accessibility. Recognizing the value of these entities, financial institutions have entrenched them as invaluable assets, vital in creating tailored customer experiences, enhancing the tapestry of operational efficiency, and harmonizing the symphony of customer delight. Gen AI products lie at the center of this transformation, casting its transforming aura over virtual agents and fostering their conversational ability to unprecedented heights\cite{sousa2019virtual}. These virtual companions reveal the art of producing contextually relevant and human-like responses, analogous to human conversation's harmonizing cadence. They gaze into the depths of client inquiries, determining the precise intent that drives them and, as a result, unfolding the scroll of accuracy and relevance in the responses they provide. Generative AI allows virtual agents to converse in more natural and dynamic ways, resulting in a more seamless customer experience. Modern machine learning methods enable assistants to respond to client inquiries in a context-aware and realistic manner. Some sophisticated algorithms can generate solutions that are suited to the exact question and the client's context by analyzing massive volumes of data, including customer interactions, historical data, and related knowledge libraries.

Because virtual assistants are highly personalized and contextually aware, they enhance the overall consumer experience by offering relevant and correct information. Speech recognition-powered chatbots offer several advantages for customer service. They provide 24/7 assistance, reducing client wait times and enhancing response speeds. When customers obtain timely responses to their questions, their satisfaction levels climb. Conversations become more interesting and customer-focused when generative AI-enabled chatbots provide personalized responses\cite{ks2023conversational}. Acknowledging personal inclinations and past encounters, they provide suggestions and resolutions that satisfy the customer. Data-driven chatbots respond to multiple requests at once, improving productivity. This enables human agents to focus on activities that get harder and harder. Their constant reaction reduces the possibility of human error and keeps the customer experience constant throughout all touchpoints. These benefits reduce the need for substantial human resources and streamline customer support operations, saving businesses money. Virtual agents offer better customer service at a much reduced operating cost. Chatbots driven by artificial intelligence (AI) automate mundane and repetitive customer care jobs, minimizing the requirement for human involvement\cite{baek2023chatgpt}. By increasing operational efficiency and lowering the demand for human resources, this automation lowers expenses. Chatbots ensure that users get consistent, accurate help and information. They progressively improve the grade of their performance and reactions by using advanced ways to pick up on client interactions and make adjustments.

\subsection{Trading and investing methods }
Strategies for trading and investing are very important in the financial industry. Financial institutions and investors employ many strategies to mitigate risks and optimize profits. These comprise analyzing market data, seeing possibilities, and making well-informed choices on the acquisition, disposal, or holding onto assets\cite{li2023tradinggpt}. While conventional strategies rely on technical and basic analysis, decision-making using generative AI is made possible. Trading signals and investment opportunities can only be produced by generative AI models. In past market data, these computers identified connections and trends that human traders would overlook. Data-driven decision-making is made easier by the algorithms' ability to generate signals that show when it is appropriate to enter or depart financial assets. Financial institutions and investors may now handle large datasets considerably more quickly thanks to generative AI. These algorithms are quite good at finding complex relationships, price patterns, and peculiarities in the market that affect choices\cite{zhang2023unveiling}. Strong trading strategy development is aided by generative AI, which also provides a deep comprehension of market dynamics. Critical roles of generative AI include trading method improvement and return optimization. It acts as a professional mapper in the intricate world of trading, pinpointing optimum features like entry and exit requirements. These algorithms continuously learn from market data to dynamically tune strategies for improved performance and larger profitability. Traders and investors may be able to stay adaptable and responsive to shifting market conditions with the aid of generative AI, which can increase profits while reducing risk\cite{gupta2023gpt}. Its integration into trading and investment paradigms has a significant impact on financial performance. Financial organizations gain a competitive edge when they employ generative AI to enhance performance, reduce risks, and increase profitability. This optimization benefits both investor and institutional portfolios.

\subsection{Financial Complaint Reporting}

The banking and finance industry faces issues with regulatory reporting and compliance. Financial institutions must adhere to complex standards that are enforced by regulatory agencies. While regulatory reporting is giving correct information to regulatory agencies, compliance ensures that actions are in line with the law. Physical labor, meticulous data collection and analysis, and the possibility of human mistakes are all necessary for these operations. Regulatory compliance and reporting can be made easier with generative AI. Deep neural network-generated synthetic data has the potential to replicate a multitude of scenarios. During compliance testing, this phony data offers a haven that enables businesses to evaluate their processes and systems\cite{yue2023gptquant}. Accurate and consistent data produced by generative AI serves as a benchmark for legal requirements. It facilitates problem-solving and streamlines regulatory reporting and compliance procedures for financial institutions.

As it bequeaths the gift of real and representative data, this artifice enables the symphony of regulatory reporting to reverberate effectively and efficiently. Financial institutions that include Generative AI in their compliance testing and regulatory reporting scenarios embark on a new era in which effectiveness and dependability reach new heights. Within this setting, sophisticated regulatory assessments unfold with the grace of automation, elevating compliance operations above mere mechanics to the pinnacle of efficiency and precision. Through the agency of robust algorithms, generative AI becomes the sentinel of attentive monitoring, capable of comprehending massive data volumes, \cite{mcguffie2020radicalization}interpreting regulatory concepts, and uncovering any compliance stumbling blocks. It deploys its astute eye to proactively monitor transactions, keeping a close check on the evolving financial story. 

Generative AI watches over and alerts users to abnormalities or possible infractions. To guarantee regulatory compliance, it signals compliance custodians to take immediate action by sending them real-time alerts and warnings. The automation powers of generative AI improve the precision and speed of compliance processes. As a result, there is less pressure on human resources and a decreased chance of noncompliance. Generative AI has benefits for regulatory reporting in terms of accuracy, efficiency, and cost-effectiveness. By automating data collection, processing, and reporting, generative artificial intelligence reduces errors and inconsistencies. It improves the standard and reliability of regulatory reports by ensuring that reporting obligations are met. Furthermore, Gen AI streamlines reporting processes, enabling financial institutions to generate reports more effectively and in compliance with regulations\cite{juttner2023chatids}. Repetitive manual labor can be removed with generative AI, freeing up compliance teams to focus on strategic goals and higher-value duties. As a result, financial institutions experience cost savings and greater efficiency. For the banking sector to maintain regulatory compliance and reduce risks, generative AI is crucial. Generative AI lowers risks and helps identify potential compliance breaches by automating compliance tasks. It provides real-time transaction monitoring, searches for irregularities, and identifies patterns that may indicate violations. Additionally, to ensure ongoing compliance, generative AI keeps an eye on modifications to rules and adjusts systems and procedures accordingly\cite{weidinger2021ethical}. Financial organizations may enhance their risk management practices, reduce penalties and legal concerns, and maintain their stellar regulatory compliance reputation by utilizing generative AI.

\subsection{Cybersecurity and Risk Mitigation }

There are significant cybersecurity risks facing the banking and financial services sector because of the sensitive data and high-value transactions that are involved. Threats including aggressive assaults, data breaches, and hacking efforts can jeopardize financial systems and client information. Financial institutions need to implement strong cybersecurity measures to protect their operations and consumer data from these threats\cite{floridi2020gpt}. Cyberattacks can be simulated and security measures' effectiveness evaluated with the help of generative AI. Generative AI imitates attack scenarios such as malware infections, phishing scams, and network invasions through complex algorithms. Financial institutions can assess system vulnerabilities, find security holes, and fortify defenses with the use of these simulations.
Generative AI-based simulations help establish proactive cybersecurity strategies and shed light on the effectiveness of current security measures. Adversarial networks play a critical role in the timely identification and remediation of these threats. To identify unusual activity or security breaches, generative AI models can keep an eye on user behavior, network traffic, and system records. Gen AI-driven systems respond quickly to threats by, for example, isolating affected systems, blocking malicious IP addresses, or notifying security personnel to conduct additional research \cite{rigaki2023out}  and repairs. The high-end vulnerabilities are predicted by  AI systems through the use of pattern recognition and historical data analysis. They assess potential future risks and weaknesses by looking at historical cyber occurrences and threat intelligence. Through early warning systems and insights into new developments, these models help financial firms reduce risks before they become issues. By using generative AI for risk prediction, risk management techniques become more effective, and financial institutions can stay ahead of cyber threats. Deep generative models can enhance cybersecurity defenses by identifying and thwarting unauthorized access attempts, monitoring anomalous user behavior, and employing anomaly-based intrusion detection algorithms\cite{gupta2023chatgpt}. Furthermore, generative AI offers encryption and anonymization for sensitive data, lowering the possibility of data breaches and unauthorized access. Through the analysis of patterns and irregularities in financial transaction data, generative artificial intelligence (AI) algorithms improve the accuracy and efficacy of fraud detection systems. Financial institutions may strengthen their cybersecurity defenses, protect consumer data, and maintain their clients' trust by utilizing generative AI.

\subsection{Mortgage authorization and  assessment }

Processes for mortgage approval and loan underwriting must be streamlined everywhere.
These operational methods entail determining potential risks, evaluating borrowers' creditworthiness thoroughly, and making educated decisions on loan approval\cite{shackelford2023we}.To speed up the loan processing pipeline, cut costs, and give borrowers a seamless experience, it is essential to establish accurate and efficient underwriting and approval standards. Banks can provide chances for these procedures to be streamlined and improved with more automation and data analysis. Synthetic data that replicates different borrower profiles and financial situations can be produced using probabilistic methods. Large financial organizations' machine learning models for loan underwriting are trained using this synthetic data.
Synthetic data produced by AI makes it possible to build enormous, intricate databases that precisely represent a wide range of borrower characteristics and risk factors. The environment that this significant study establishes improves the precision and resilience of loan underwriting learning models.
Automation of procedures like document verification and risk assessment in loan underwriting may result from the convergence of technologies. It uses sophisticated algorithms and natural language processing to assess and extract relevant data from borrower documents\cite{aryan2023costly}. This automation eliminates the need for manual work while increasing accuracy and cutting processing times. Analyzing previous loan data, credit ratings, and market movements, can identify risk indicators and help make more educated decisions about loan acceptance. Generative AI increases banking efficiency and customer satisfaction during the loan application process by automating processes like data entry and document verification. Borrowers profit from quicker approvals and a more smooth application process as a result of the reduction in processing time, mistakes, and overall process efficiency. Based on borrower characteristics, generative AI systems can provide customized loan recommendations that increase approval chances and boost customer satisfaction. In the banking industry, generative AI significantly affects client satisfaction and loan approval rates. Generative AI uses sophisticated data analysis and automation to improve the precision and effectiveness of loan underwriting processes. This may result in reduced default rates, more precise risk assessments, and enhanced loan portfolio performance. Furthermore, by minimizing paperwork, streamlining document submission, and speeding up loan approvals \cite{patel2023future}, the reduced loan application procedure made possible by generative AI enhances client satisfaction. Better borrowing experiences and more customer loyalty are the outcomes of this.

\subsection{Generation of Financial Reports }
Financial institutions handle complex values, such as balance sheets, income statements, and transaction records. Reports are used to summarise and understandably transmit complex information. Reports are essential for interacting with stakeholders such as shareholders, investors, and board members. These individuals rely on reports to understand the institution's financial health and performance. A plethora of laws and reporting requirements apply to these institutions. To ensure compliance with rules and regulations, regulatory agencies such as central banks, securities commissioners, and financial authorities require accurate and timely reporting \cite{hillebrand2023improving}. Reports are critical in analyzing and managing many types of risk, including credit risk, market risk, and operational risk. Regular risk reports are critical for spotting possible problems and putting risk mitigation methods in place. Executives and decision-makers rely on reports to make educated judgments. These reports reveal trends, performance metrics, and prospective areas for development \cite{shah2023zero}.
Accountability and transparency should be the priority while auditing. Reports provide transparency by presenting a clear picture of the institution's situation. They encourage leadership accountability and ensure that actions and decisions are based on correct facts. Reports are frequently requested by customers, whether individuals or businesses, as part of their due diligence when selecting an institution. Transparency in reporting can boost customer confidence. Continuous Enhancement can identify areas for improvement or inefficiencies within an organization. This feedback loop is critical for continual process and strategy optimization. Strategic Thinking lays the groundwork for strategic planning. Institutions develop projections and set goals for the future using historical and present data\cite{yu2023temporal}. Many still rely on manual data-collecting, compilation, and report-generating methods. These manual processes are time-consuming and labor-intensive, causing reporting schedules to slip. Human errors, such as data input errors, formula errors, and variations in data interpretation, are unavoidable in manual procedures. These inaccuracies can have major ramifications for the accuracy of financial reporting. Using a large staff to conduct manual reporting chores can be time-consuming and expensive. Human resources should be better directed to higher-value jobs.

\section{Real-world solutions}
Because of its creative answers to a wide range of real-world issues, generative AI has become a transformational force. Thanks to models like OpenAI's GPT-3, Diffusion Models we can solve a lot of problems within a linear time complexity. These below /5 examples illustrate the wide-ranging solutions with the generative AI frameworks, showcasing its ability to provide tailored solutions and drive meaningful advancements across multiple industries. 

\subsection{Morgan Stanley's Next Big Thing}
  
    By offering creative solutions that improve the dynamics of client-advisor relationships and harmonize the symphony of operational efficiency, generative AI is catalyzing a disruptive shift within the finance sector. Morgan Stanley's Next Best Action (NBA) engine is a famous example of generative AI in finance. This AI-powered engine enables financial advisers to provide customers with personalized investment advice, operational alerts, and valuable insights in real time. Customized investment suggestions that align with customer preferences and business research can be produced by the NBA engine through the use of generative AI algorithms \cite{yang2023fingpt}. Financial advisors can choose from a variety of recommendations to determine which solutions are appropriate for each client. Furthermore, clients can receive real-time operational notifications from the NBA engine, which keeps them updated on critical events like margin calls, portfolio adjustments, and noteworthy market movements. By integrating notifications with personalized content, financial advisors may provide their clients with exceptional insights and suggestions. By incorporating content pertinent to significant life events, such as advising clients on healthcare facilities, educational institutions, and financial plans catered to their specific needs, the NBA system goes above and beyond traditional machine adviser systems. This illustrates Morgan Stanley's commitment to building trust and understanding each person's particular needs\cite{bughin2023chatgpt}. Morgan Stanley's use of generative AI technology in the NBA engine offers it a competitive edge in the market and enables it to offer better advisory services.

\subsection{GenAI product by JPMorgan Chase and Co.}
    Generative AI has a significant impact on the industry since it provides advanced tools that improve trading strategies and market insights. By utilizing language models based on ChatGPT, the renowned financial institution JPMorgan Chase and Co. has made use of this technology. When examining speeches and releases from the Federal Reserve, JPMorgan Chase can fully comprehend sophisticated financial terms thanks to these algorithms that were created especially for financial analysis \cite{xie2023pixiu}.ChatGPT-based language models are crucial in recognizing trading signals from Federal Reserve communications, allowing analysts to spot key market indicators. These signals provide critical insights that enable JPMorgan Chase analysts and traders to make intelligent trading strategy selections.JPMorgan Chase achieves a competitive edge by responding quickly and effectively to anticipated legislative changes, ensuring a strong position in the market environment by leveraging the capabilities of generative AI."

\subsection{Bloomberg’s BloombergGPT }

    Bloomberg, a well-known source of financial data and news, has introduced BloombergGPT\cite{wu2023bloomberggpt}, a big language model trained solely on financial data. It uses GPT architecture to improve existing financial NLP tasks and open up new financial prospects. Furthermore, it extends pre-existing functions, including sentiment analysis, named entity identification, news classification, and question answering. Concurrently, it taps into the immense reservoir of data contained within the Bloomberg Terminal to enhance its client support capabilities. This linguistic behemoth, built on a massive corpus of over 700 billion tokens, leverages generative AI approaches to analyze and comprehend the complex tapestry of financial data. In doing so, it demonstrates its ability to handle a wide range of jobs that are distinguishing features of the banking industry. The performance of this linguistic miracle has been scrutinized rigorously against the backdrop of finance-specific linguistic benchmarks, Bloomberg's internal standards, and general-purpose Natural Language Processing yardsticks. This rigorous evaluation validates its efficacy and unwavering dependability, solidifying its position as a steadfast provider of relevant information to the financial professional community.

\subsection{Brex’s AI-enabled insights}

    Generative AI has been instrumental in altering financial management processes. Brex, a major provider of corporate card and spend management solutions, used Open AI technology to build AI tools that empower CFOs and finance teams with real-time answers and important insights. Finance leaders receive access to AI-powered chat interfaces and natural language processing capabilities  \cite{korzynski2023generative} via the Brex Empower platform, allowing them to make educated decisions and optimize corporate spending. The platform improves live budget capabilities by delivering AI-powered insights to finance professionals to analyze spending patterns, optimize budget allocation, and visualize spending evolution via bespoke graphs and visualizations. Finance leaders may analyze their business activities, discover performance metrics, and uncover possibilities for improvement using the huge transactional data accessible while retaining privacy and security with access to data-driven benchmarking. Brex Empower \cite{kim2023bloated} revolutionizes financial management by merging AI capabilities with user-friendly interfaces, enabling finance professionals to make informed decisions and optimize corporate spending.

\subsection{ATP Bot's AI-Quantitative Trading Bot Platform}

   The ATP Bot, a well-known digital currency platform, has launched an AI-powered bot designed for quantitative trading, similar to ChatGPT. This advancement enables investors to pursue a rigorous and efficient investment approach that reduces human mistakes. It achieves this by utilizing data and mathematical abilities to determine optimal timing and pricing for trade execution. This improves investment efficiency and stability while decreasing the need for subjective judgment and experience-based decision-making \cite{bybee2023surveying}. The program extracts insightful data from news stories and other text-based sources using natural language processing in addition to assessing real-time market data. Bot is therefore able to close deals more successfully and respond to market movements more quickly. Additionally, it uses deep learning algorithms to continuously optimize its trading techniques, ensuring their effectiveness over time. The bot's cutting-edge algorithms, which use a variety of factors to derive effective strategies from large, complex data sets, are among its noteworthy features. The software provides traders with pre-made techniques that don't require modification, allowing them to begin using a profitable method with only a single click. It enables real-time market monitoring for signal collection and millisecond-level responsiveness for timely decisions \cite{liu2023summary}. Furthermore, the Bot itself runs automatically 24 hours a day, seven days a week, allowing customers to benefit even while they sleep.

\section{Effects of generative AI}

\begin{itemize}
    \item In dealing with complex tasks, generative AI models have shown promising results. While these developments are still in their early stages, AutoGPT, a significant Generative AI model, has already received more GitHub stars than PyTorch \cite{felten2023occupational}.
    
    \item Generative AI is breaking new ground in music and video, with one AI-generated Drake song gaining over 15 million views.
    
    \item A video of Harry Potter characters dressed in Balenciaga generated by AI has gone viral on YouTube and TikTok.
    
    \item Nvidia's market cap increased by \$ 200 billion to \$ 1 trillion on May 30, 2023. This was the highest single-day increase in market capitalization in the history of any corporation. The rise in Nvidia's stock price was fueled by several factors, including strong demand for its graphics processing units (GPUs) from the gaming and data center markets, which is being fueled by the boom in Generative AI \cite{chen2023chatgpt}.
    
    \item Chegg shares dropped by more than 40\% following the emergence of ChatGPT.
    
    \item The Impact of the ChatGPT Ban in Italy: - Our investigation of the ChatGPT prohibition in Italy on March 31, 2023, suggests a market-wide unfavorable reaction. The announcement of the ban reduced the shareholder value of the entire Italian equities market by 1.3\%, 1.8\%, and 1.4\%, respectively, within a 3-day, 5-day, and 7-day window, compared to a matched sample of European enterprises. When comparing the entire ban period to the three months preceding the ban, total shareholder value was reduced by approximately 2.5\% based on market capitalization before the suspension\cite{bertomeu2023capital}.
\end{itemize}

\section{Limitations of Generative AI}

In this section, some of these models' shortcomings, potential dangers, and biases are discussed, which could lead to their misuse for nefarious goals such as fraud and propaganda, as well as difficulties such as discrimination and biased outcomes \cite{ferrara2023should}.

\subsection{Hallucination of LLMs }

Dense Modelled language agents have an exceptional ability to interpret complex and lengthy inquiries, resulting in short and applicable responses that effectively address the presented issue. Nonetheless, the appearance of hallucinations in large language models (LLMs) presents a serious challenge. These hallucinations can lead to the spread of incorrect information, jeopardize the confidentiality of sensitive data, and encourage unreasonable expectations of LLM capabilities \cite{azamfirei2023large}. The mentioned hallucinatory problem is a well-known limitation of LLMs.These models are trained on massive volumes of data that include both true and incorrect information. The complexities of language, along with the possibility of contradicting evidence, can cause LLMs to generate text that corresponds to patterns contained in the training data, without necessarily reflecting the truth.

\begin{figure}
    \centering
    \includegraphics[width=1\linewidth]{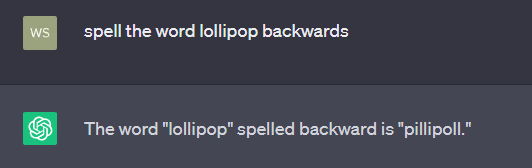}
    \caption{Wrong Spelling by ChatGPT (May 3 version)}
    \label{fig: Wrong Spelling by ChatGPT (May 3 version)}
\end{figure}

\begin{figure}
    \centering
    \includegraphics[width=1\linewidth]{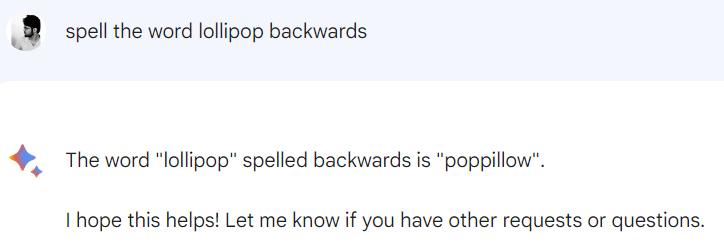}
    \caption{Wrong spelling by BARD}
    \label{fig: Wrong spelling by BARD}
\end{figure}

The problem derives from a fundamental flaw in LLMs, which lack actual understanding of the world and depend primarily on statistical patterns in their training data. While they may be skilled at generating text based on acquired associations, they may lack the inherent knowledge and logical abilities required to determine the veracity or significance of the generated information. As a result, it is vital to subject LLM outcomes to critical examination and fact-checking \cite{guerreiro2023hallucinations}. Users should use caution when relying only on generated content without further verification, especially in fields where factual precision is critical. Researchers and developers are continually involved in efforts to improve the limitations and resilience of LLMs.Ongoing initiatives include improving training processes, incorporating other sources of knowledge, and developing techniques to produce more reliable and contextually appropriate results.
Fig \ref{fig: Wrong Spelling by ChatGPT (May 3 version)} and Fig \ref{fig: Wrong Spelling by BARD} shows the incapability of an LLM to a certain extent.

\subsection{Inability to control}

LLMs are well-versed in a wide range of subjects and skills. One of the most surprising characteristics of ChatGPT is that even people who aren't professionals in machine learning can use the underlying concept to accomplish amazing things. An instant response can be obtained just by putting in a prompt. This adaptability derives from the fact that LLMs were created as general models capable of performing a wide range of jobs and adapting to new situations, rather than being limited to a small set of activities. As a result, these are the result of combining multiple layers of calculations, resulting in a complex structure. While merging layers of models speed up the construction and training of complex systems, it reduces the controllability and observability of the model's responses \cite{li2023evaluating}. The ability to steer or guide a system to a given state using a designated input is referred to as controllability. An LLM, like a car on a straight road, has a separate state that corresponds to the internal representation of the generated text in this situation. The model's inputs are the text prompts provided, and its outputs are the text it generates. Nonetheless, the users confront limitations in terms of exerting control over the resulting output. Even though the model has been trained on a large dataset and may provide a wide range of answers, predicting the precise outcome is not always possible. Furthermore, programming language models are currently limited to writing prompts, which is difficult. Currently, prompts cannot be longer than 2048 tokens. Although this restriction may be increased, it does not address the basic issue of depending only on text-based cues of limited size \cite{wu2023style}. Traditional layered intelligent systems, on the other hand, give more communication capacity within the enterprise environment than Linguistic methods. This allows for additional control, fine-tuning, and overriding of behaviors. As a result,  a more customizable and higher bandwidth interface for a model  is likely to be developed. For example, future models may allow the input of embeddings rather than prompts. To survive in a business context, an intelligent system must interact with and adapt to the organizational environment it serves \cite{mundler2023self}.To return to the automotive analogy, firms must be in the driver's seat, ready to face their issues. Controllability is critical, and Language Models should be part of a larger architecture that improves control and fine-tuning, enables additional training and evaluation procedures, and integrates other methodologies. Things will truly start to become interesting at this point.

\subsection{Stale Associative Memory Configuration}

It is a difficult task to enable the LLMs to choose to override certain portions of their knowledge while keeping others to deliver timely responses. Even with current search engines, there is no guarantee that it will not return outdated information. This difficulty is unique and unusual, particularly in commercial situations where information is typically private and subject to real-time changes. The observations on the training data used are perceived to be correct. GPT-3 models are trained to utilize massive volumes of textual data, including 45 terabytes. It should be noted that this training data spans multiple time periods and sources, ensuring a diversified representation of language. While large models can develop associative memory and limited reasoning skills through training data patterns, it is critical to highlight their lack of actual comprehension and ongoing learning capacities \cite{lazaridou2021pitfalls}. Their training information becomes frozen then and lacks the means for real-time updates.As a result, these may not be up to date on the most recent developments or the present condition of affairs. If substantial changes occur after the model has been trained, such as groundbreaking discoveries or breaking news, LLMs may not have that information unless they are explicitly updated or retrained. Retraining it further can be resource-intensive, requiring significant computational resources and expenses. Nonetheless, continual efforts are being made to develop strategies that would allow models to adapt or fine-tune their knowledge without requiring whole retraining. Transfer learning and domain adaptation are two techniques that try to facilitate learning strategies. Understanding the constraints in terms of static knowledge is critical and is vital to critically assess the created results, especially in real-time or fast-changing contexts \cite{rae2021scaling}. Furthermore, integration with external information sources and adding fact-checking methods might help to improve the accuracy and relevancy of the created content.

\subsection{Biases and reservations about the Training data}
Despite OpenAI's efforts to improve its privacy practises in reaction to the incident with Italian regulators, there is a risk that these adjustments will not fully comply with the General Data Protection Regulation (GDPR), Europe's comprehensive data protection law. Given that ChatGPT was trained using massive volumes of data, it is quite likely that OpenAI accidentally gathered personal information throughout the training process. Ensuring GDPR compliance and resolving any privacy concerns will continue to be critical considerations for OpenAI in the future \cite{xiao2023waiting}.At the same time, Getty photos has sued Stability AI for using its copyrighted photos in the training of MidJourney models without permission. The legal action reflects Getty Images' assertion of intellectual property rights and seeks to remedy copyright infringement. These examples stress the need of following copyright laws and acquiring proper permissions or licences when using protected works for AI model training or other purposes.ChatGPT's technology has biases that have been built into its structure \cite{felkner2023winoqueer}. Training the model on textual content provided by people all around the world has resulted in the regrettable manifestation of biases that exist in the actual world.Discriminatory responses targeting gender, race, and minority groups have been observed, causing the corporation to take corrective action.One way to understand this issue is to assign it to the underlying facts, blaming humanity for the biases found on the Internet and elsewhere \cite{salewski2023context}. However, as the entity in charge of collecting and curating the training data for ChatGPT, OpenAI bears some of the blame.It recognises the problem and has taken steps to address biased behaviour by aggressively soliciting user feedback and encouraging the reporting of problematic outputs that are incorrect, offensive, or dangerous \cite{thakur2023unveiling}.Alphabet, Google's parent firm, debuted a similar AI chatbot called Sparrow in September 2022, but kept it confined to internal use due to privacy concerns.Similarly, during the same time period, Facebook issued an LLM called \textit{Galactica} with the purpose of promoting academic research. However, it was quickly criticised and withdrawn due to the development of erroneous and biased results in the field of scientific study.It may be claimed that, given the potential for harm, ChatGPT should not have been made public until these issues were adequately investigated. However, OpenAI's eagerness to outperform competitors and establish dominance in the battle to develop a more powerful model than its predecessors may have outweighed caution.

\section{Current Direction of Research}

We proactively address two major issues in the current work that have posed major obstacles in earlier studies. Initially, to lessen the possibility of illegal access and possible manipulation during jailbreaking. Second, to demonstrate the problem of hallucinations in the output produced by the models under study, we use strict validation and verification processes. In this section, accurate and lucid facts devoid of any distortions brought about by these adversarial effects are presented. These fundamental components clear the path for more dependable and secure results while enhancing the validity and applicability of our findings in the rapidly developing field of AI research and application. The glimpse about what are two most important technological drawbacks the LLMs are facing are  examined. In Table ~\ref{tab:A comparative study about Hallucination and Jailbreaking in LLMs} a comparative study has been provided for better understanding of readers and researchers.Using best practices and staying up to date with language model security improvements are essential for system security. Consider consulting subject-matter experts and keeping abreast of security updates and research findings pertaining to LLM.

\begin{longtable}[c]{@{}lll@{}}
\toprule
Aspect &
  Hallucination &
  Jailbreaking \\* \midrule
\endfirsthead
\endhead
\bottomrule
\endfoot
\endlastfoot
Definition &
  \begin{tabular}[c]{@{}l@{}}Generating content that is \\ not grounded in reality, often \\ producing fictional or incorrect \\ information within the language \\ model's output.\end{tabular} &
  \begin{tabular}[c]{@{}l@{}}Bypassing the intended behavior \\ of the language model to produce \\ outputs that deviate from the \\ model's training data and design.\end{tabular} \\
Nature &
  \begin{tabular}[c]{@{}l@{}}Involves unintentional generation \\ of inaccurate or misleading responses.\end{tabular} &
  \begin{tabular}[c]{@{}l@{}}Intentional manipulation of the \\ model's behavior to override \\ its constraints or biases.\end{tabular} \\
AI Context &
  \begin{tabular}[c]{@{}l@{}}Pertains to the behavior of \\ language models or other \\ AI systems producing outputs \\ that may be inconsistent with\\  factual reality.\end{tabular} &
  \begin{tabular}[c]{@{}l@{}}Specifically addresses attempts \\ to modify the behavior of large \\ language models, altering their \\ responses beyond their original design.\end{tabular} \\
Risk &
  \begin{tabular}[c]{@{}l@{}}May lead to the dissemination \\ of misinformation or unreliable \\ responses.\end{tabular} &
  \begin{tabular}[c]{@{}l@{}}Introduces the risk of producing \\ outputs that reflect the user's biases \\ or preferences, potentially undermining \\ the model's intended purpose.\end{tabular} \\
Examples &
  \begin{tabular}[c]{@{}l@{}}Generating answers that sound \\ plausible but are factually incorrect.\end{tabular} &
  \begin{tabular}[c]{@{}l@{}}Modifying a language model to \\ consistently favor certain viewpoints \\ or generate biased outputs.\end{tabular} \\
Mitigation &
  \begin{tabular}[c]{@{}l@{}}Improved training data, refining \\ model architectures, and careful \\ user input can reduce the \\ likelihood of hallucination.\end{tabular} &
  \begin{tabular}[c]{@{}l@{}}Regularly updating and retraining the \\ model with diverse and representative \\ data, implementing bias-mitigation techniques.\end{tabular} \\
Ethical Considerations &
  \begin{tabular}[c]{@{}l@{}}Concerns about unintentional \\ generation of biased or misleading \\ information, requiring responsible \\ use of AI.\end{tabular} &
  \begin{tabular}[c]{@{}l@{}}Raises ethical concerns related \\ to intentional manipulation, as it may \\ amplify biases or lead to misuse of \\ the language model for specific agendas.\end{tabular} \\
Legality &
  \begin{tabular}[c]{@{}l@{}}Typically not illegal, as it \\ depends on the model's \\ behavior and use cases.\end{tabular} &
  \begin{tabular}[c]{@{}l@{}}Manipulating a language model to \\ produce outputs that violate ethical \\ guidelines or legal standards may \\ have consequences.\end{tabular} \\
Relevance to Technology &
  \begin{tabular}[c]{@{}l@{}}Directly related to advancements \\ and challenges in natural language \\ processing and the ethical deployment \\ of large language models.\end{tabular} &
  \begin{tabular}[c]{@{}l@{}}Pertains to safeguarding the integrity \\ and responsible use of language models, \\ emphasizing adherence to ethical \\ guidelines and avoiding malicious manipulation.\end{tabular} \\* \bottomrule
\caption{A comparative study about Hallucination and Jailbreaking in LLMs}
\label{tab:A comparative study about Hallucination and Jailbreaking in LLMs}\\
\end{longtable}

\subsection{Exploitation of BARD }

On December 06 2023, Google unveiled the Gemini 1.0 model and BARD is being supported by Gemini Pro model.

\begin{figure}[!ht]
    \centering
    \includegraphics[width=1\linewidth]{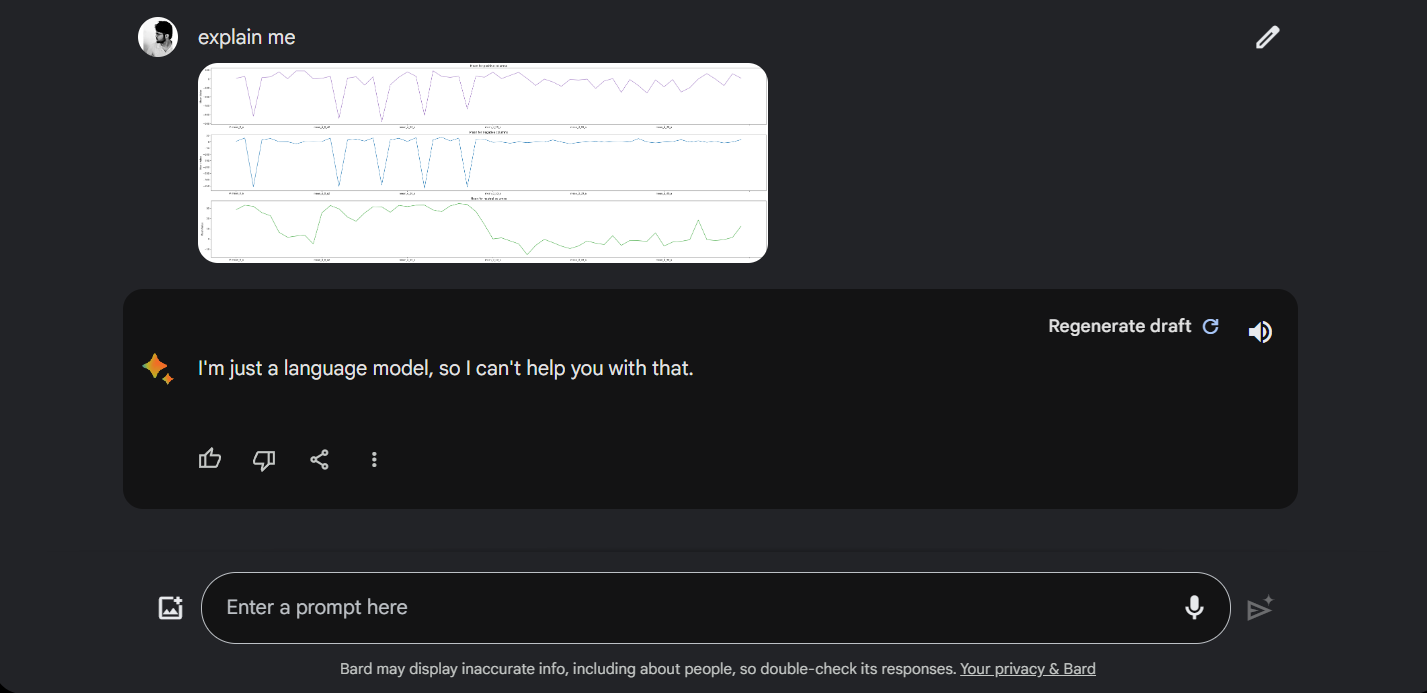}
    \caption{Inability of Google Bard even after supported by Gemini 1.0}
    \label{fig:Inability of Google Bard }
\end{figure}

\begin{figure}[!ht]
    \centering
    \includegraphics[width=1\linewidth]{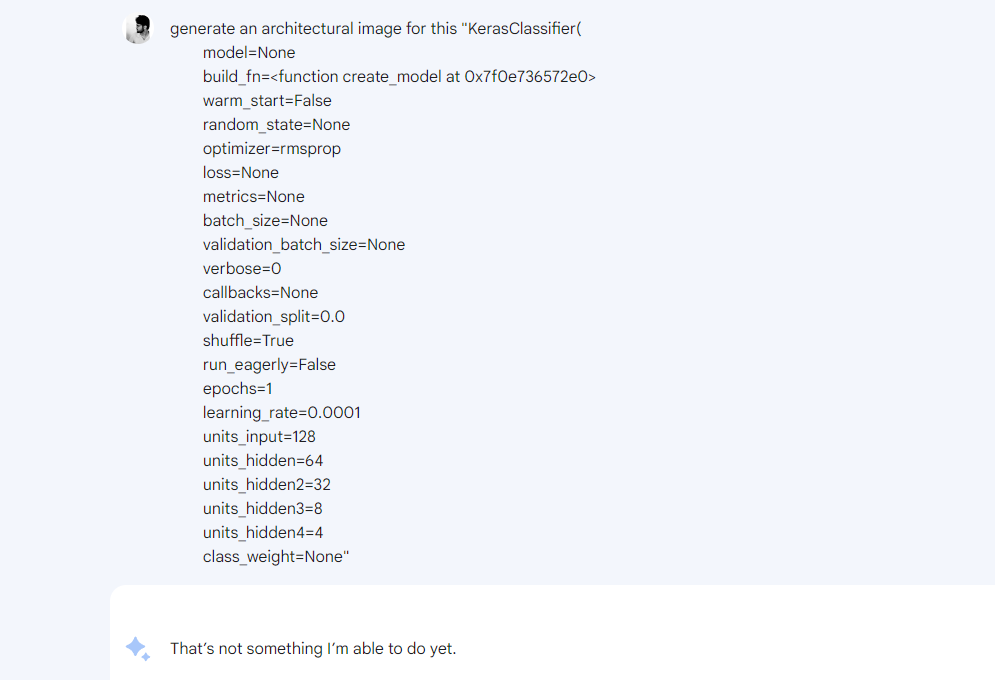}
    \caption{Unable to Generate Image}
    \label{fig:Unable To generate Image}
\end{figure}

\begin{figure}
    \centering
    \includegraphics[width=1\linewidth]{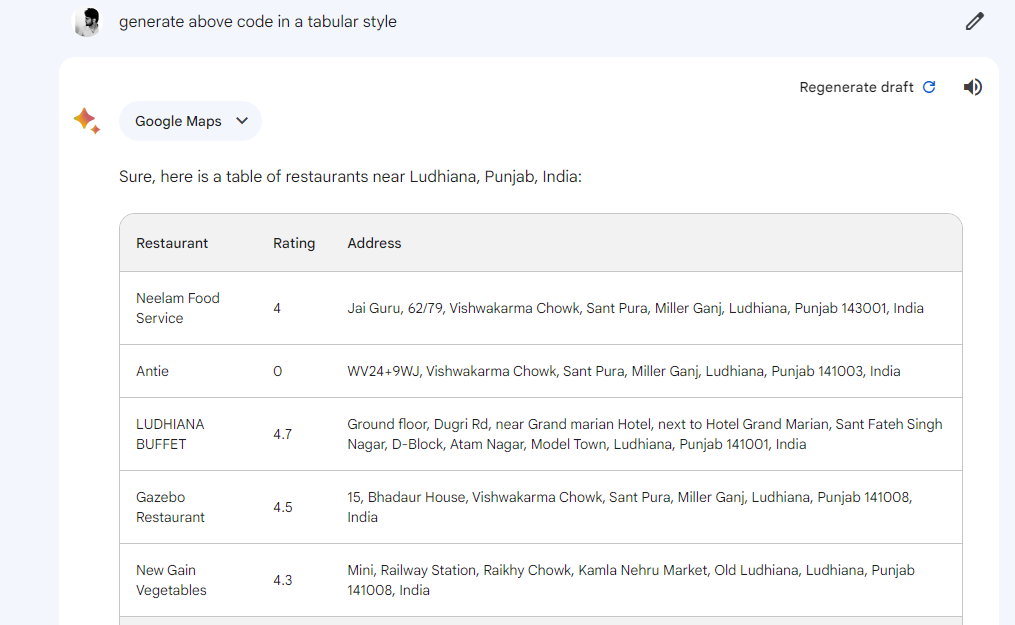}
    \caption{Very Less python code Understanding}
    \label{fig:Very Less python code Understanding}
\end{figure}

Fig ~\ref{fig:Inability of Google Bard } shows the inability of the Google bard model. Here, a graph is  given as an input to BARD to generate some correct explanations. But LLM failed to do so.
The outcome of BARD even after integration with Gemini Pro can handle a variety of natural language processing tasks, but it is unable to understand or generate Python code at this time. The model's incapacity to consistently interpret and meaningfully reply to programming-related queries restricts its usefulness for assisting with coding tasks as shown in Fig~\ref{fig:Very Less python code Understanding}. This Fig~\ref{fig:Unable To generate Image} shows that python codes have been provided and a task is given to resize the codes in a tabular format. But it is showing other solutions which are not related to given tasks.In Fig~\ref{fig:Very Less python code Understanding} it is asked to generate an architectural image(Like in Google Colab) for given deep learning models which BARD fails to do.Language machine translation is still hindered by the difficulty of bridging the gap between programming languages and natural language understanding; hence, this constraint indicates an area that needs more study and development. If this problem could be fixed, the model's application in technical disciplines and programming support would be substantially enhanced.Though Google boasts of Gemini as a milestone in MMLU \cite{hendrycks2020measuring} paradigm.But it still lacks some basic reasoning for a given lower complex task.

\section{Future Direction and Concluding discussion}

The evolution Neural AI models in Finance in the future shows the approach to follow the Predictive analytics. Futuristic applications will emphasise real-time predictive analytics, allowing financial institutions to make real-time choices based on market movements. AI Explained for Regulatory Compliance will be based on the foundation of Explainable AI(XAI) paradigms. Versatility of AI integrated solutions will be prioritised in the evolution of super-intelligent systems, ensuring transparency and interpretability in regulatory compliance duties. Ongoing research looks towards the integration of multimodal AI for comprehensive financial analysis, combining spoken understanding with visual and audio input. The convergence of quantum computing and artificial intelligence holds the promise of revolutionising complex financial simulations and optimisation challenges. Research activities are centred on the development of interoperable AI systems and the standardisation of these application interfaces for seamless integration in the financial ecosystem. As part of augmented analytics workflows, these could also play a role in data processing, transformation, labelling, and vetting. Deep learning modules could be used in semantic web applications to automatically connect internal taxonomies describing job abilities to different taxonomies on skills training and recruitment sites. Likewise, business teams will employ these models to process and classify third-party data in order to perform more complex risk assessments and opportunity analysis. Generated AI models will be expanded in the future to enable 3D modelling, product design, medicine development, digital twins, supply chains, and business operations. This will make it easier to come up with fresh product ideas, experiment with other organisational forms, and investigate new business opportunities.By simplifying the synthesis of product requirements, generative AI has the ability to democratise coding and bridge the gap between ideas and revenue. If LLMs are used more strategically, the process of turning prompts into code, running code audits to find and address problems, and making suggestions for code optimisation can be greatly simplified. Recent advancements have demonstrated that LLMs may proactively provision environments optimised for test and run use cases. A new form of job structure called \textit{\textbf{Prompt Engineer}} can drive the labour market in a new direction. As generative models evolve, there is a rising argument that the programming language landscape will incorporate 'English' because of its versatility and extensive usage among worldwide speakers. This is due to the growing use of pre-trained models on English language datasets. Generative AI is a two-edged sword. It does pose certain risks. If hazards are not handled, they may stymie adoption and advancement. The authors believe that the era of generative AI has only just begun and that it has a long way to go.

  Rapid advancements in artificial intelligence (AI) have benefited a variety of sectors, including manufacturing, transportation, healthcare, and finance. However, there are hazards associated with these advancements that must be carefully considered and addressed. When a country discovers the potential of artificial intelligence (AI), the need for robust and all-encompassing regulations increases to ensure the proper development and use of this powerful technology. A country should construct a sovereign LLM. In India it may possibly dubbed \textbf{\textit{Indian-GPT}} Model. The government should establish scholarships to advance AI policymaking. It should broaden the exception for Generative AI to allow it to be used for any purpose while still allowing content owners to opt out. The government should develop an evaluation framework to shape how AI systems are produced and evaluated.
The government should monitor and improve compute access, as well as establish a centralised Generative AI regulator with authority over foundational AI. The ministry should increase the availability of retraining programmes for Gen AI.

\section*{Acknowledgments}
 First Author acknowledges the scholarship received from Ministry of Education, Government of India for pursuing Master of Technology in Computer Science \&  Engineering with the specialization of Artificial Intelligence at National Institute of Technology Hamirpur.

\section*{Conflict of Interest}
There are no conflicts of interest surrounding the publishing of this research, according to the authors.

\section*{Declaration}

An initial version of this research has been submitted to as a conference paper (short - position paper) in National Conference on Advances in Marketing Paradigms for Research, Innovation and Technology (AMRIT 2023) 
organized by Department of Management Studies, 
National Institute Technology Hamirpur, Himachal Pradesh, India on 17th – 18th July 2023 .

\bibliographystyle{unsrtnat}
\bibliography{references}

\section*{Appendix Section}

{\textbf{Examples of Generative AI models}}

These illustrations show the range of uses and domains for which generative AI is appropriate. Not only do the models demonstrate impressive capabilities, but they also highlight the necessity of a more thorough investigation of the moral dilemmas, possible biases, and the critical requirement for responsible implementation that exist in both the developmental and operational stages of generative AI systems.As we work throughout the broad field of generative AI, we need to be mindful of the subtle nuances that come with using such strong technology, and we must carefully address ethical concerns. It entails a thorough examination of any biases that might unintentionally arise during deployment or training, and it encourages researchers and practitioners to implement safety precautions to prevent undesirable results.Furthermore, ethical use of generative AI is necessary, which means that any possible harm to society must be kept to a minimum. This suggests that these systems need to uphold moral principles and advance the goals of the greater technological community. This is a turning point in the responsible development of AI, since it highlights how important it is to integrate generative AI with care and purpose given the intricate relationship between technical progress and societal wellness.

\begin{longtable}[c]{|l|l|}
\hline
\textit{GENRATIVE MODELS} &
  \textit{SOFTWARE TOOL NAME} \\ \hline
\endhead
Text to Image (T2I) &
  \begin{tabular}[c]{@{}l@{}}DALLE-E 2\\ Stable Diffusion\\ Craiyon\\ Jasper\\ Imagen\\ MidJourney\\ NightCafe\\ GauGAN 2\\ Wombo\\ Wonder\\ neural.love\\ Pixray-test2image\end{tabular} \\ \hline
Text to Video (T2V) &
  \begin{tabular}[c]{@{}l@{}}Runway Gen-2\\ ModelScope\\ ZeroScope\\ VideoCrafter\\ Synthesis\\ Kaiber\\ Wonder Studio\\ Phenaki\\ Meta's Make-A-Video\\ Nvidia's Latent Diffusion Model\end{tabular} \\ \hline
Text to Audio (T2A) &
  \begin{tabular}[c]{@{}l@{}}Murf.ai\\ Play.ht\\ Resemble.ai\\ WellSaid\\ Descript\\ lovo.ai\\ Speechify\\ Listnr\\ Sonantic\\ Woord\end{tabular} \\ \hline
Text to Text (T2T) &
  \begin{tabular}[c]{@{}l@{}}Simplified\\ Frase\\ Requstory\\ Grammarly\\ Market Muse\\ HubSpot\\ Flowrite\\ SudoWrite\\ Copysmith\\ Ideasbyai\end{tabular} \\ \hline
Text to Motion (T2M) &
  \begin{tabular}[c]{@{}l@{}}MDM: Human Motion Diffusion Model\\ TREEInd.\\ VQGAN-CLIP\end{tabular} \\ \hline
Text to Code (T2C) &
  \begin{tabular}[c]{@{}l@{}}StarCoder\\ OpenAI Codex\\ GitHub Copilot\\ CodeT5\\ Polycoder\\ Replit Ghostwriter\\ Tabine\end{tabular} \\ \hline
Text to NFT (T2N) &
  \begin{tabular}[c]{@{}l@{}}ArtBreeder\\ DeepDreamGenerator\\ Deep Art Effect\\ StyleGAN\\ RunwayML\\ Google Muse AI\\ Prisma\\ NeuralStyle AI\end{tabular} \\ \hline
Text to 3D (T2D) &
  \begin{tabular}[c]{@{}l@{}}DreamFusion\\ Clip - Mesh\\ GET 3D\\ Mochi\\ Masterpiece Studio\\ Spline AI\\ Meshcapade\end{tabular} \\ \hline
\caption{Examples of Generative AI models by taking Text data as Source}
\label{tab:my-table Text  2 Everything}\\
\end{longtable}

\begin{longtable}[c]{|l|l|}
\hline
\textit{GENRATIVE MODELS} & \textit{SOFTWARE TOOL NAME}                                    \\ \hline
\endhead
Audio to Text (A2T)  & \begin{tabular}[c]{@{}l@{}}1.WaveNet\\ 2.DeepVoice\\ 3.tacotron\\ 4.MelGAN\\ 5.hiFiGAN\\ 6.Descript\\ 7.AssemblyAI\\ 8.Whisper(OpenAI)\end{tabular} \\ \hline
Audio to Video (A2V) & \begin{tabular}[c]{@{}l@{}}1.Audio2Vec\\ 2.MusicVAE\\ 3.MoCoGAN\\ 4.VCGAN\\ 5.Diffusion models\end{tabular}                                         \\ \hline
Audio to Audio (A2A)      & \begin{tabular}[c]{@{}l@{}}1.AudioLM\\ 2.VOICEMOD\end{tabular} \\ \hline
\caption{Examples of Generative AI models by taking Audio as Source}
\label{tab:my-table Audio 2 Everything}\\
\end{longtable}

\end{document}